\documentclass{aa}
\usepackage[varg]{txfonts}
\usepackage{graphicx}	
\usepackage{hyperref}

\hypersetup{
    colorlinks=true,
    linkcolor=blue,
    citecolor=blue,
    filecolor=black,
    urlcolor=blue,
}

\title{Sensitivity of solar wind mass flux to coronal temperature}
\author{
	D. Stansby \inst{\ref{inst1}}\thanks{d.stansby@ucl.ac.uk} \and
	L. Ber\v{c}i\v{c}  \inst{\ref{inst2},\ref{inst3}}\and
	L. Matteini  \inst{\ref{inst4}}\and
	C. J. Owen  \inst{\ref{inst1}} \and
	R. J. French  \inst{\ref{inst1}}\and
	D. Baker  \inst{\ref{inst1}}\and
	S. T. Badman \inst{\ref{inst5},\ref{inst6}}
}
\institute{
	Mullard Space Science Laboratory, University College London, Holmbury St. Mary, Surrey RH5 6NT, UK \label{inst1} \and
	LESIA, Observatoire de Paris, PSL Research University, CNRS, UPMC University Paris 6, University Paris-Diderot, Meudon, France \label{inst2} \and
	Physics and Astronomy Department, University of Florence, Sesto Fiorentino, Italy \label{inst3} \and
	Department of Physics, Imperial College London, London, SW7 2AZ, UK \label{inst4} \and
	Physics Department, University of California, Berkeley, CA 94720-7300, USA \label{inst5} \and
	Space Sciences Laboratory, University of California, Berkeley, CA 94720-7450, USA \label{inst6}
}

\begin{document}
\abstract{
Solar wind models predict that the mass flux carried away from the Sun in the solar wind should be extremely sensitive to the temperature in the corona, where the solar wind is accelerated. We perform a direct test of this prediction in coronal holes and active regions, using a combination of in-situ and remote sensing observations. For coronal holes, a 50\% increase in temperature from 0.8 MK to 1.2 MK is associated with a tripling of the coronal mass flux. At temperatures over 2 MK, within active regions, this trend is maintained, with a four-fold increase in temperature corresponding to a 200-fold increase in coronal mass flux.
}

\maketitle

\section{Introduction}
The Sun continuously loses mass through the solar wind. Although the rate of this mass loss small at 
2 $\times$ 10$^{-14}$ M$_{\odot}$ yr$^{-1}$ \citep{Cohen2011}, it plays an important role in transporting angular momentum away from the Sun, controlling the rate at which it spins down \citep{Weber1967, Li1999}.

The solar wind mass flux can be predicted with simple hydrodynamic models, where the number density is supplied as a lower boundary condition in the corona and an equation of state relating the temperature and density is assumed \citep{Parker1958, Parker1960}. For a given base number density, under spherical expansion the mass flux depends on the temperature profile inside the sonic point via.
\begin{equation}
	n_{\odot} v_{\odot}  = n_{\odot} c_{\odot} \left ( \frac{r_{c}}{r_{\odot}} \right )^{2} \frac{c_{\odot}}{c_{c}} \exp \left [- \frac{1}{2} \int_{r_{\odot}}^{r_{c}}  r_{c} \frac{w^{2}}{ c^{2} ( r ) r^{2}} dr \right ],
	\label{eq:mass flux}
\end{equation}
where $c^{2} \propto T$ is the thermal speed, $w$ is the solar escape velocity, $r$ is radial distance from the centre of the Sun, $n$ is number density, $c$ subscripts are values evaluated at the critical sonic point (where $v_{c} = c_{c}$), and $\odot$ subscripts are values evaluated at the solar surface \citep[][eq. 25]{Parker1964}. An increase in $T$ results in a decrease of the integral, which in turn results in the increase in the mass flux. The exponential dependence of mass flux on temperature means that under spherical expansion, small variations in coronal temperature should result in large variations in mass flux. Such large variations are not seen in the solar wind mass flux at 1 AU however \citep{Leer1982, Withbroe1989, Goldstein1996}.

The resolution of this apparent inconsistency involves two competing effects cancelling each other out: areas of strong magnetic field in the corona undergo stronger heating driving increased mass fluxes, but stronger magnetic fields also undergo more super-radial expansion, resulting in a diluting of the mass over a larger area \citep{Wang2010}. Correcting for radial magnetic field expansion and calculating near-Sun coronal (as opposed to solar wind) mass fluxes can be done routinely using magnetic field models, and it has been shown that the mass flux in the corona spans many orders of magnitude \citep{Wang1995, Wang2010, Schwadron2008}. Correlating these large variations with temperature changes is challenging however, as coronal temperatures are hard to reliably measure remotely \citep[e.g.][]{Habbal1993, Esser1995}, and in-situ measurements of solar wind temperatures at 1 AU have been significantly distorted from their coronal values \citep[e.g.][]{Marsch1983, Stansby2019a, Maksimovic2020}.

In this paper we perform such a direct comparison using spectroscopic observations of two active regions, and a newly proposed in-situ proxy for coronal temperature in three coronal hole streams \citep{Bercic2020}. Section \ref{sec:methods} briefly discusses the methods used to calculate coronal mass fluxes, and sections \ref{sec:chs} and \ref{sec:ars} present the solar wind streams and temperature measurements for coronal holes and active regions respectively. Section \ref{sec:results} presents the main results, showing that a four-fold increase in coronal temperature is associated with a 200-fold increase in coronal mass flux. The results are discussed and put in the context of other studies in section \ref{sec:discussion}, with conclusions provided in section \ref{sec:conclusions}.

\section{Methods}
\label{sec:methods}
In order to compare mass fluxes over a wide range of coronal temperatures, data from both coronal holes and active regions were used. To infer coronal mass fluxes, in-situ measurements of the solar wind mass flux were scaled back to their coronal values using the frozen in theorem \citep[e.g.][]{Wang2010}:
\begin{equation}
	n_{\odot}v_{\odot} = n_{sw} v_{sw} \frac{\left | \mathbf{B}_{\odot} \right |}{\left | \mathbf{B}_{sw} \right |},
	\label{eq:coronal flux}
\end{equation}
where $n$ is the number density, $v$ the radial velocity and $\mathbf{B}$ the magnetic field. A ${sw}$ subscript denotes a quantity measured in the solar wind and a ${\odot}$ subscript denotes a quantity evaluated at the base of the corona.

Measurement of coronal temperatures were done using different methods for coronal holes and active regions.
For coronal holes it is hard to reliably determine coronal temperatures with remote sensing data \citep[e.g.][]{Habbal1993, Wendeln2018}, so a new method was used, which provides a local in-situ proxy for the coronal temperature. For active regions, hotter temperatures (and therefore higher ultra-violet emission intensities) allowed the use of remote sensing spectroscopy to estimate the coronal temperature.

\section{Data}
\label{sec:data}
In this section the choice of discrete solar wind streams is discussed, along with the data used to estimate coronal mass fluxes and temperatures for each source type. A summary of the data collected for each stream, and the various data sources, is given in table \ref{tab:summary}.

\begin{table*}
\caption{Summary of the parameters and data sources of the five solar wind streams. The median value is given where a range of values is measured within each stream.}
\label{tab:summary}
\centering
\begin{tabular}{c c c c c c c}
\hline\hline
Stream	& $\left | \mathbf{B} \right |_{\odot}$  (G)	&$\left |\mathbf{B} \right |_{sw} r^{2} / r_{\odot}^{2}  (G)$	& $n_{sw}v_{sw}r^{2} $ ($10^{35}$sr$^{-1}$s$^{-1}$	)	& $n_{\odot}v_{\odot}r_{\odot}^{2} $ ($10^{35}$sr$^{-1}$s$^{-1}$	)	& $T_{corona}$ (MK)	& Solar wind, $B_{\odot}$, $T_{corona}$ data	\\
\hline
S1 		& 1.54 							& 1.10										& 0.66										& 0.92		& 1.13 					& PSP, HMI, PSP  \\
S2		& 1.54							& 0.99										& 0.40										& 0.63		& 0.93					& PSP, HMI, PSP \\
S3		& 1.61							& 1.61										& 0.38										& 0.38		& 0.79					& PSP, HMI, PSP \\
AR1		& 19.2							& 2.91										& 0.42										& 2.66		& 1.86					& WIND, MDI, EIS	  \\
AR2		&  255							& 1.47										& 0.43										& 69.2		& 2.28					& WIND, MDI, EIS	 \\
\hline
\end{tabular}
\end{table*}
\subsection{Coronal holes}
\label{sec:chs}
\subsubsection{Choice of streams}
Data taken by \emph{Parker Solar Probe} \citep[PSP,][]{Fox2016} during its first perihelion were used to compare the properties of three coronal hole streams. 1 minute averaged magnetic field data were taken from the FIELDS instrument suite \citep{Bale2016} level 2 data. Solar wind proton core density and velocity data were taken from the \emph{Solar Wind Electron Alpha and Proton} suite \citep[SWEAP,][]{Kasper2016} level 3 data, with 1 minute mean values taken to align the plasma data with the magnetic field data.

\begin{figure}
	\includegraphics[width=0.48\columnwidth]{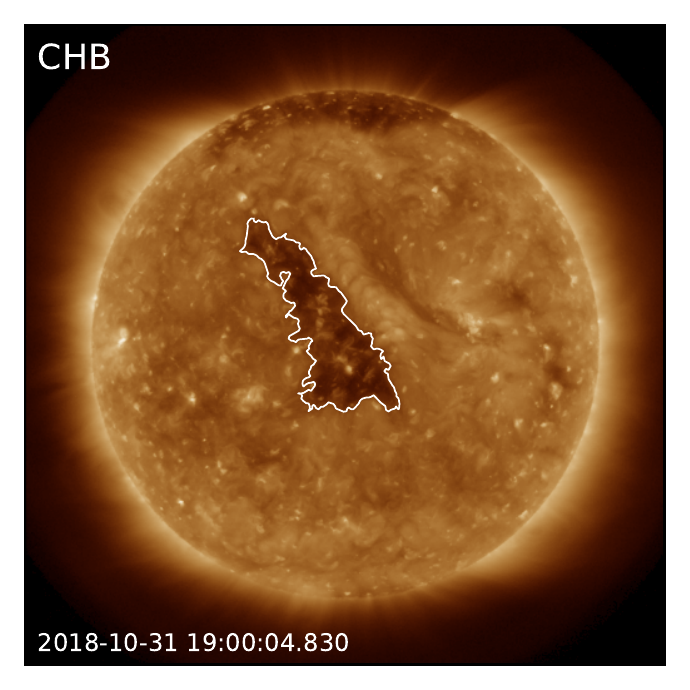}
	\includegraphics[width=0.48\columnwidth]{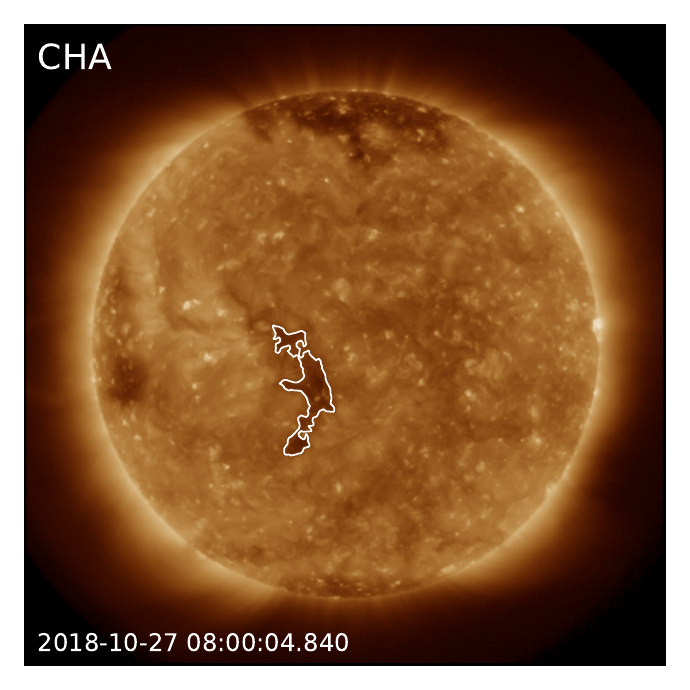}
	\includegraphics[width=\columnwidth]{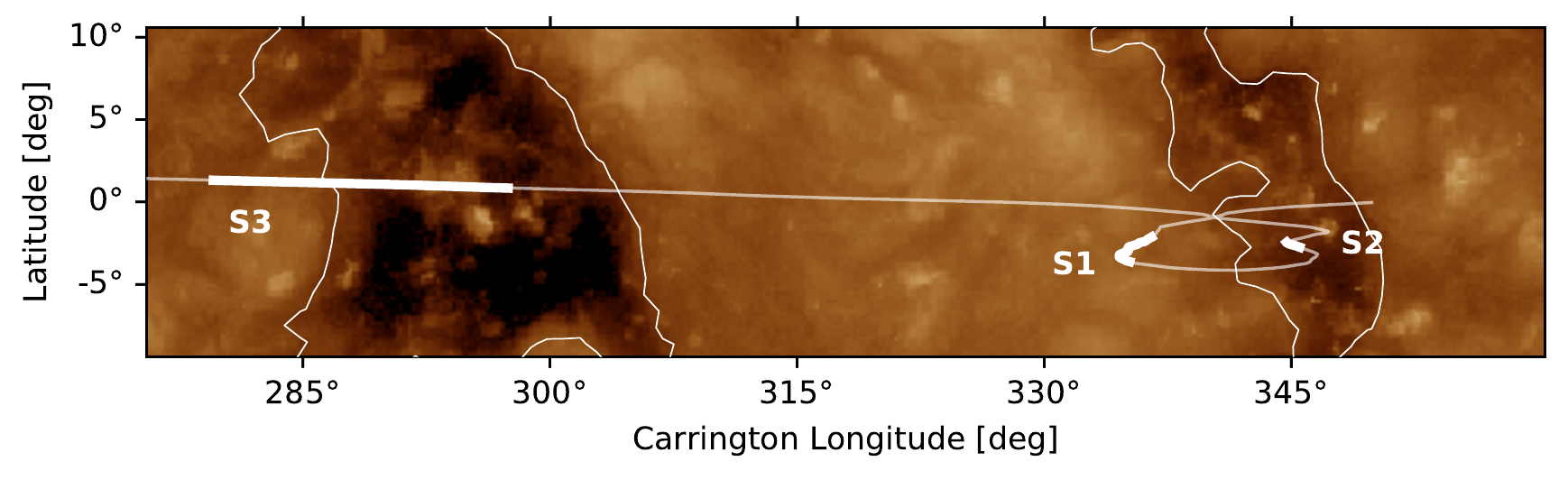}
	\caption{The top panel shows two coronal holes sampled by PSP, with contours showing the identified coronal hole with an intensity threshold at 50~DNs$^{-1}$. The y-axis is aligned with solar north in both images, and the colourscale is clipped at 3000 DNs$^{-1}$. The bottom panel shows a Carrington map of the same two coronal holes, with the white line showing PSP's trajectory ballistically backmapped to 2.5$r_{\odot}$. Labelled areas of the trajectory are the in-situ data intervals selected for analysis.}
	\label{fig:CH}
\end{figure}
Magnetic mapping during the first perihelion pass shows PSP was initially connected to a small equatorial coronal hole, and subsequently connected to a second larger equatorial coronal hole \citep{Badman2020}. These are respectively labelled `CHA' and `CHB', and Fig.~\ref{fig:CH} shows images of them, taken by the \emph{Atmospheric Imaging Assembly} \citep[AIA,][]{Lemen2012} on board the \emph{Solar Dynamics Observatory} \citep[SDO,][]{Pesnell2012}. The top left panel shows the large coronal hole and top right panel shows the smaller coronal hole. Both were isolated using an intensity contour at 50~DN~s$^{-1}$.

The bottom panel of Fig.~\ref{fig:CH} shows a synoptic map with the trajectory of PSP ballistically backmapped to 2.5R$_{\odot}$. In this co-rotating Carrington coordinate system the spacecraft moved from right to left with time, performing a loop at closest approach. The highlighted portions of the trajectory indicate the three intervals selected for further analysis, labelled S\{1,2,3\}. The first two were located on either side of the perihelion loop over the small coronal hole, and the third interval was located over the large coronal hole.

\begin{figure}
	\includegraphics[width=\columnwidth]{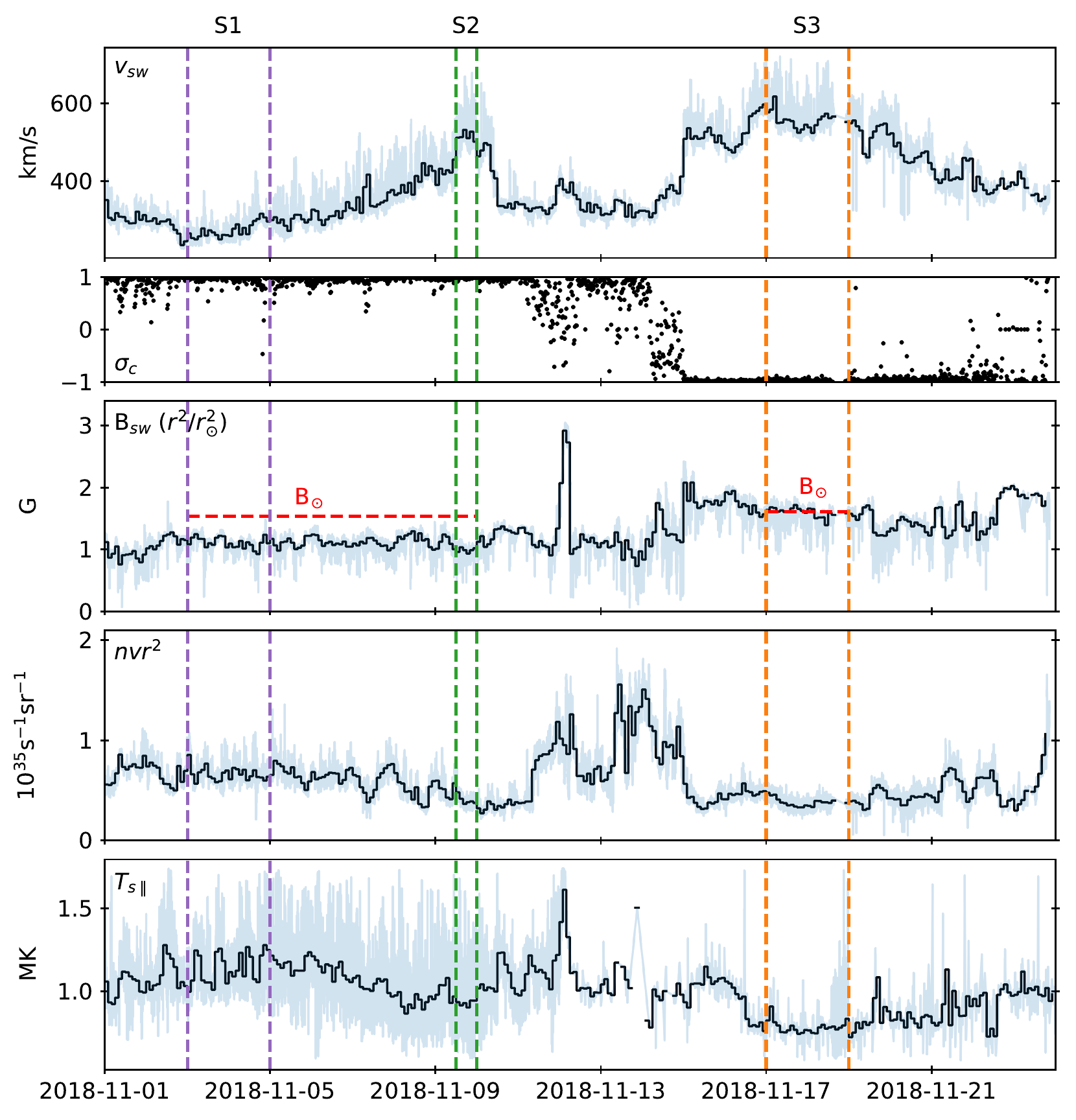}
	\caption{PSP measurements during perihelion 1, with vertical dashed lines bounding the three distinct streams discussed in the text. Light grey lines are 1 minute measurements and solid black lines 2 hour averages. Top panel shows the solar wind velocity, and second panel shows cross helicity calculated on a 20-minute scale. The third panel shows the in-situ magnetic field, scaled by $(r / r_{\odot})^{2}$ to make it directly comparable to coronal magnetic field strengths. Horizontal dashed lines show the corresponding magnetic field strength in the source coronal holes. The fourth panel shows solar wind mass flux, and the fifth panel shows the parallel electron strahl temperature.}
	\label{fig:tseries}
\end{figure}
Fig.~\ref{fig:tseries} shows an overview of solar wind parameters measured by PSP, with the three streams indicated with coloured bands. The top panel shows the solar wind speed. The streams marked in Figs.~\ref{fig:CH} and \ref{fig:tseries} were selected to have a relatively constant velocity, avoiding any stream interaction regions \citep[e.g.][]{Perrone2018}. Stream S1 was the lowest velocity interval during perihelion. Although it had very slow speeds ($\approx 250$ km s$^{-1}$), S1 was strongly Alfv\'enic, as shown the second panel which plots cross helicity \citep[calculated as in][on a 20 minute timescale]{Stansby2019d}. This adds confidence that it originated in a coronal hole \citep[e.g.][]{DAmicis2018, Stansby2020}. It originated away from the centre CHA (see Fig.~\ref{fig:CH}), consistent with a large expansion factor and therefore slower speed. Stream S2 had a higher speed at around 500 km s$^{-1}$. This stream was also highly Alfv\'enic and originated in the same coronal hole as S1. This part of the trajectory was directly over the centre of the small coronal hole, likely consistent with a smaller expansion factor and therefore faster speed. Stream S3 was another high speed stream at the end of the interval. This was a typical fast solar wind stream, highly Alfv\'enic and with speeds around 500 km s$^{-1}$. It originated from the large coronal hole (CHB) shown in the top left panel of Fig~\ref{fig:CH}, and was measured by PSP during the outbound leg of its orbit, at 0.3 -- 0.5 AU.

\subsubsection{Coronal magnetic field}
\label{sec:ch mag}
The magnetic field strength in each coronal hole was calculated from photospheric magnetograms measured by the \emph{Heliospheric Magnetic Imager} \citep[HMI,][]{Scherrer2012} on SDO. Coronal hole boundaries were taken from intensity thresholds at 50 DN~s$^{-1}$ on AIA 193~\AA~images (shown in Fig.~\ref{fig:CH}), and the mean magnetic field (corrected for projection effects) within these boundaries calculated using the method of \cite{Hofmeister2017}. Taking a single average value assumes a spatially isotropic magnetic field strength with each coronal hole, which is a good assumption at the base of the corona where the plasma beta is $\ll 1$ \citep{Peter2006}.

The third panel of Fig.~\ref{fig:tseries} shows coronal magnetic field strengths in context with the in-situ data. Dashed red lines show the coronal hole magnetic field strengths, and the solar wind data are scaled by $(r/r_{\odot})^{2}$. This allows the expansion factor to be visualised, defined as $f =  ( B_{\odot}r_{\odot}^{2} ) / ( B_{sw}r^{2} )$. For streams S1 and S2 the photospheric magnetic field was larger than the scaled in-situ field, giving an expansion factor $> 1$. In contrast stream S3 had almost identical photospheric and scaled in-situ field strengths, giving an expansion factor of unity.

\subsubsection{Coronal temperature}
\label{sec:ch temp}
The fourth panel of Fig.~\ref{fig:tseries} shows the parallel strahl electron temperature \citep{Bercic2020}. This is defined as the gradient of high energy electrons (the strahl) in velocity space. Under adiabatic expansion, and due to the low collisionality of the high energy electrons, the strahl temperature should be conserved from when the corona was last collisionally dominated to where it is measured the solar wind, giving a proxy for the coronal temperature \citep{Bercic2020}. Although there is a large scatter between individual measurements, there are clear trends visible across the whole interval. During S1 the temperature was relatively high, and then gradually declined as the solar wind speed increases until S2. During S3 the measurements were sparser, but on average this interval had a lower temperatures than the previous intervals.

\subsection{Active regions}
\label{sec:ars}

\subsubsection{Choice of streams}
Two active regions were analysed, previously studied both remotely and in-situ by \cite{vanDriel-Gesztelyi2012} and \cite{Stansby2020d} respectively. These studies used magnetic modelling and ballistic backmapping to identify the in-situ solar wind intervals at 1 AU corresponding to each active region.  In-situ data were measured by the \emph{Solar Wind Experiment} \citep[SWE][]{Ogilvie1995} and the \emph{Magnetic Field Investigation} \citep[MFI][]{Lepping1995} on board WIND, from 2008 January 12 14:00 UT to 2008 January 13 12:00 UT for AR1 and 2013 January 24 00:00 UT to 2013 January 25 00:00 UT for AR2.
\begin{figure}
	\includegraphics[width=\columnwidth]{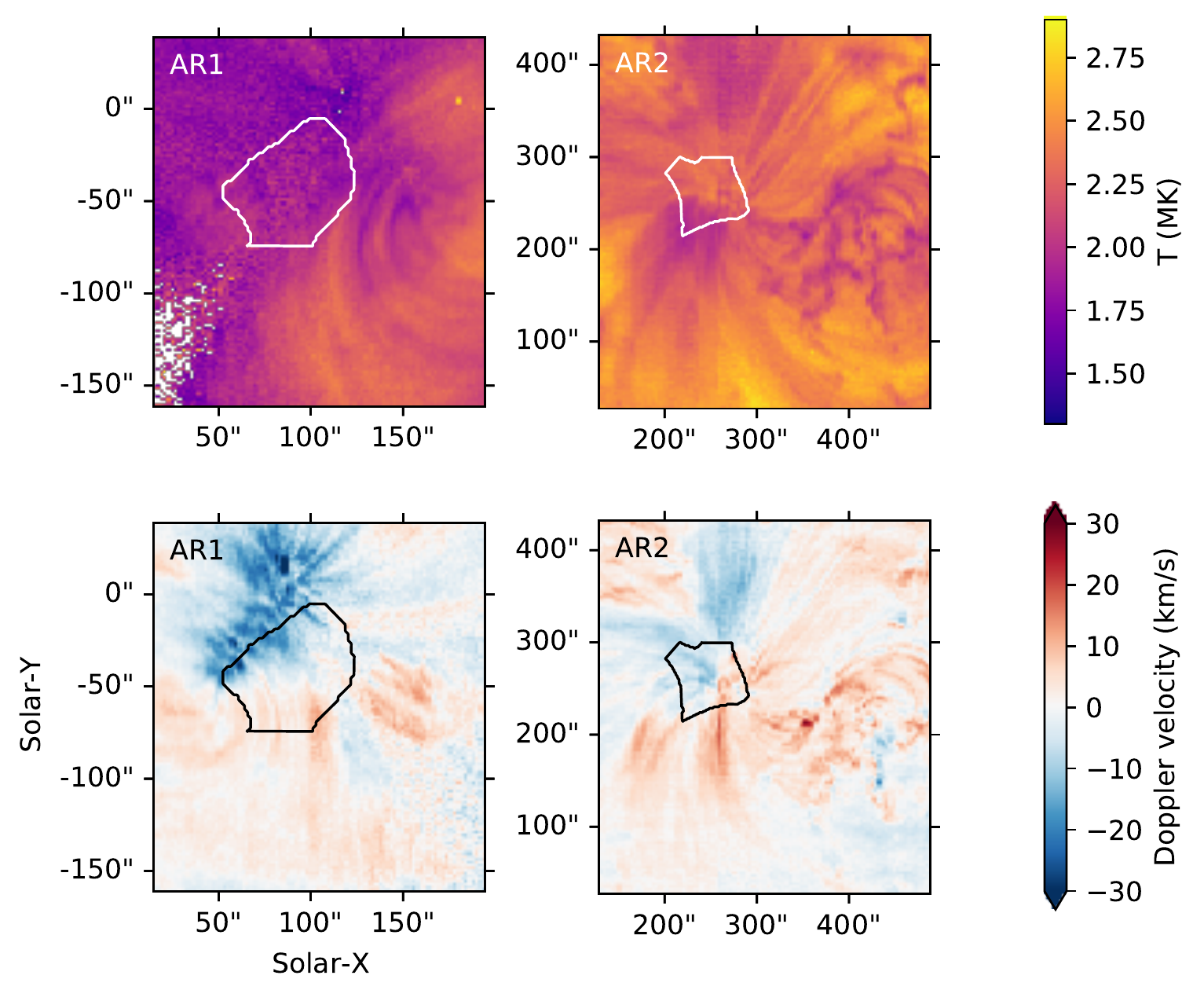}
	\caption{Observations of two active regions used in this study. Top panels show Doppler velocity, with negative (blue) values indicating flows away from the Sun. Bottom panels show plasma temperature. The black and white contours outline regions of open magnetic field. Note that the spatial scales are different for the two regions, with the area of the open field region $\sim$4 times larger in AR2 than in AR1.}
	\label{fig:eis}
\end{figure}

\subsubsection{Coronal magnetic field}
To isolate the areas in the corona responsible for feeding the solar wind, open-closed field maps were calculated around each active region, by tracing field lines through a potential field source surface \citep[PFSS,][]{Altschuler1969, Schatten1969} model. The models were calculated from \emph{Global Oscillation Network Group} \citep[GONG,][]{Harvey1996, Plowman2020} synoptic photospheric magnetic field maps, using the \texttt{pfsspy} software package \citep{Stansby2020e}, with a source surface radius at 2.5R$_{\odot}$.

The open/closed field contour for each active region is shown over-plotted in Fig.~\ref{fig:eis}. To measure the coronal magnetic field, the open field regions were isolated on high resolution line of sight field maps, from the \emph{Michelson Doppler Imager} \citep[MDI,][]{Scherrer1995} for AR1 and from HMI for AR2. The average photospheric field within the open field contour was calculated as in section \ref{sec:ch mag}.

\subsubsection{Coronal temperature}
\label{sec:ar temp}
Spatially resolved spectrographic data from \emph{Hinode}/EIS \citep{Kosugi2007, Culhane2007} were used to measure electron temperatures in the active regions. The EIS data were prepped and fitted using the \emph{SolarSoftWare} \texttt{eis\_prep} and \texttt{eis\_auto\_fit} routines. The Fe XIII 202.04 to Fe XII 195.11 \AA~lines observed by EIS are a temperature-sensitive line pair, with good sensitivity at active region temperatures  \citep[][section 11.1]{DelZanna2018}. Using the theoretical ratio of these lines computed in CHIANTI v8 \citep{DelZanna2015}, temperature maps were calculated for the two active regions. These electron temperature maps are shown in the top two panels of Fig.~\ref{fig:eis}. As an additional check on whether coronal material was flowing into the solar wind \citep[e.g.][]{Harra2008, Marsch2008}, line of sight Doppler velocity maps were also calculated for the Fe XIII 202.04 \AA~line\footnote{With the reference wavelength (ie.~zero velocity point) set assuming zero average shift over the entire map.}, shown in the bottom panels of Fig.~\ref{fig:eis}. The distribution of temperatures within each active region were taken from pixels within the open field contour, that had negative Doppler velocities (ie. material was flowing away from the Sun).

\section{Results}
\label{sec:results}
Using equation \ref{eq:coronal flux}, point by point measurements of solar wind mass flux divided by magnetic field strength were multiplied by the average photospheric source magnetic field to give a distribution of coronal mass fluxes for each source. The distribution of coronal temperatures was taken either from the strahl parallel temperatures within coronal holes (see section \ref{sec:ch temp}) or the active region temperatures determined through spectroscopy (see section \ref{sec:ar temp}).

\begin{figure}
	\includegraphics[width=\columnwidth]{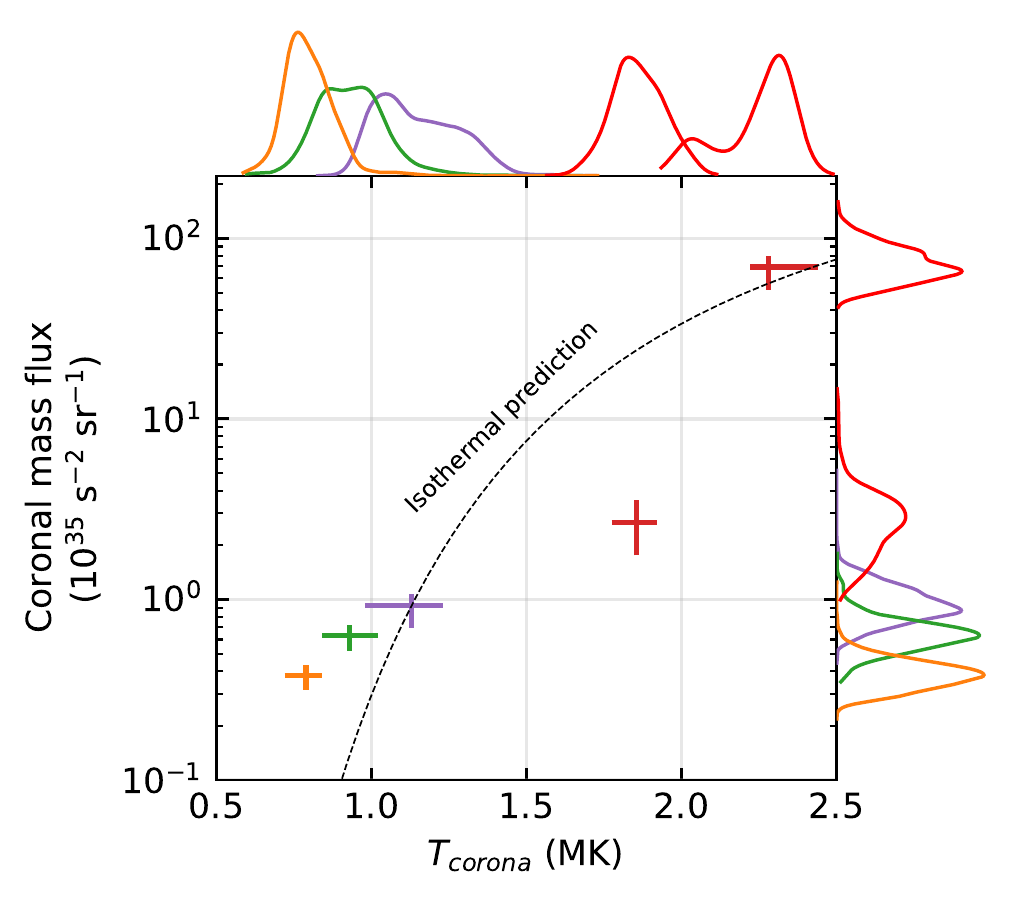}
	\caption{Joint distributions of coronal temperature and coronal mass flux for three coronal holes (lower left) and two active regions (upper right). Crosses are centred on the median values and span the 20th - 80th percentiles. Marginal 1D distributions show Gaussian kernel density estimations. The dashed line is an isothermal prediction for $ \left | \mathbf{B} \right | \propto 1/r^{2}$.}
	\label{fig:hists}
\end{figure}

Fig.~\ref{fig:hists} shows the the variation of coronal mass flux with coronal temperature across the five different streams analysed here. Within the coronal hole streams, the fastest stream had the lowest temperatures ($\approx 0.8$ MK), and the lowest mass fluxes ($\approx$ 0.3 $\times$ 10$^{35}$ s$^{-2}$ sr$^{-1}$). In contrast, the slow wind interval had temperatures around 50\% higher than this, and mass fluxes around 200\% higher. The intermediate coronal hole interval confirms this trend, lying between the other intervals in both temperature and mass flux.
The active regions both had higher temperatures, at 2 MK and 2.4 MK, and significantly higher mass fluxes, at 2 and 80 $\times$ 10$^{35}$ s$^{-2}$ sr$^{-1}$ respectively. Between the coolest coronal hole and hottest active region, an increase in temperature by a factor of around 4 results in an increase in the coronal mass flux by a factor of over 200.\footnote{We stress that this is the mass flux at the base of the corona; the solar wind mass flux does not vary by such large orders of magnitude.}

Making a quantitative comparison to theory is challenging, as even in simple fluid models the evolution of magnetic field strength and temperature as a function of height must be provided as inputs. The simplest prediction comes from a fluid model which assumes both $B \propto 1/r^{2}$ and an isothermal temperature profile, which reduces equation \ref{eq:mass flux} to \citep[see][eq. 31]{Parker1964}:
\begin{equation}
	n_{\odot}v_{\odot} = n_{\odot}c R_{c}^{2} e^{-2 \left ( R_{c} - 1 \right )},
\end{equation}
where $R_{c} = r_{c} / r_{\odot} = w^{2} / (4c^{2}) \approx (5.8~\textup{MK}) /  T$ is the sonic point normalised to the solar radius. Taking a typical observed value of $n_{\odot} = 2 \times 10^{8}$ cm$^{-3}$ \citep{DelZanna1999}, this prediction is shown as the dashed line in Fig.~\ref{fig:hists}. This model agrees qualitatively with the data, in the sense that it predicts the correct order of magnitude and large variations in the mass flux. However, the trend fails to accurately predict mass fluxes for the cool coronal holes and intermediate temperature active region. This is unsurprising, as both the magnetic field strength and temperature profiles are not constant in the corona; more accurate model assumptions need to be considered to understand if fluid models successfully predict mass loss rates.

\section{Discussion}
\label{sec:discussion}
The observation that coronal mass flux is extremely sensitive to corona temperature agrees qualitatively with fluid theories of the solar wind, which also predict the correct order of magnitude for the mass flux. To make more accurate quantitive comparisons, observed magnetic field strengths and coronal temperature profiles need to be measured. In the corona $\left | \mathbf{B} \right |$ can only be directly measured below about 1.5$r_{\odot}$, and even then measurement is limited to brighter areas away from coronal holes \citep{Wiegelmann2014, Yang2020}. This could be circumvented by density and velocity measurements \citep{Bemporad2017} or magnetic field orientation measurements \citep{Boe2020} which can indirectly infer expansion factors in the corona. Temperature profiles can be estimated using off-limb spectroscopy \citep[e.g.][]{Landi2008, Cranmer2020}, or again using density observations to indirectly infer temperatures \citep{Lemaire2016}.

In contrast to this study of individual solar wind streams, changes in mass flux and coronal temperatures can be measured over multiple 11-year solar cycles. In the minimum between cycles 23 and 24 the mass flux in polar coronal holes was lower than the minimum between cycles 22 and 23 \citep{McComas2008, McComas2013, Zerbo2015}. This reduction was accompanied by a reduction in oxygen charge state ratios, which implies a corresponding reduction in the coronal temperature \citep{Zhao2011, Schwadron2014}. Our study agrees well with, and provides a stream by stream verification of these long duration statistical variations.

The mass flux carried away from a star controls stellar spin down, with angular momentum loss rate directly proportional to mass loss rate \citep{Weber1967}. Indeed, the reduction in coronal temperatures and therefore solar wind mass flux between cycles 23 and 24 drove a similar reduction in the solar angular momentum loss \citep{Finley2019, Finley2019a}. Our results suggest that if there was a way to remotely measure the coronal temperature of the parts of stars in which stellar winds originate, it would be possible to predict the mass loss rate. Unfortunately only globally integrated observations are available for other stars, which are dominated by closed-loop emission \citep{Cohen2011, Mishra2019}. However, our observations can be used to place new constraints on the mass fluxes predicted by solar and stellar wind models \citep[e.g.][]{Johnstone2015, Usmanov2018, Shoda2020}. 
\section{Conclusions}
\label{sec:conclusions}
We have presented a comparison of solar coronal temperatures and mass fluxes, across three coronal holes and two active regions. A factor of four increase in coronal temperature results in over two orders of magnitude increase in mass flux in the solar wind sources studied, confirming that solar wind mass flux in the corona is extremely sensitive to the plasma temperature. This study provides a new insight into understanding solar mass loss via.~the solar wind, which in the future can be extended to large statistical studies and detailed theoretical comparisons.

\begin{acknowledgements}
D.S., C.J.O., and D.B. are supported by STFC grant ST/S000240/1. R.F. is supported by STFC grant ST/S50578X/1. Hinode is a Japanese mission developed and launched by ISAS/JAXA, collaborating with NAOJ as a domestic partner, NASA and UKSA as international partners. Scientific operation of the Hinode mission is conducted by the Hinode science team organised at ISAS/JAXA. This team mainly consists of scientists from institutes in the partner countries. Support for the post-launch operation is provided by JAXA and NAOJ (Japan), UKSA (U.K.), NASA, ESA, and NSC (Norway). We acknowledge the NASA Parker Solar Probe Mission and the FIELDS and SWEAP teams for use of data. We thank the anonymous referee for providing comments that helped improve presentation and interpretation of our results. Data processing was carried out with the help of astropy \citep{TheAstropyCollaboration2018}, and sunpy \citep{TheSunPyCommunity2020}. Figures were produced using Matplotlib \citep{Hunter2007} \\ \\

Code to reproduce the figures presented in this paper is available at \url{https://github.com/dstansby/publication-code}. PSP and WIND data are available from \url{https://spdf.gsfc.nasa.gov/pub/data}, GONG data from \url{https://gong2.nso.edu/oQR/zqs/}, SDO and SOHO data from \url{http://jsoc.stanford.edu/}, and EIS data from \url{http://solarb.mssl.ucl.ac.uk/SolarB/}.

\end{acknowledgements}

\bibliographystyle{aa}
\bibliography{/Users/dstansby/Dropbox/zotero_library}
\end{document}